\documentclass[%
superscriptaddress,
reprint,
amsmath,amssymb,
aps,
prl,
10pt,
twocolumn,
floats,
]{revtex4-2}

\usepackage[english]{babel}
\usepackage{graphicx}
\usepackage{dcolumn}
\usepackage{amsmath}
\usepackage{epsfig}
\usepackage{bm}
\usepackage{amssymb}
\usepackage{float}
\usepackage{natbib}
\usepackage{color}
\usepackage{soul}
\usepackage{hyperref}
\hypersetup{colorlinks=true,linkcolor=blue,citecolor=blue,urlcolor=black}

\usepackage{units}

\newcommand{\comment}[1]{\color{blue}\it \color{black} \rm}

\newcommand{\ket}[1]{|#1 \rangle}

\newcommand{\mmu}{\ensuremath{\mu}}

\begin{document}

\title{Slowing Down a Coherent Superposition of Circular Rydberg States of Strontium}

 \author{L. Lachaud}
\affiliation{Laboratoire Kastler Brossel, Coll\`ege de France,
 CNRS, ENS-Universit\'e PSL,
 Sorbonne Universit\'e, \\11, place Marcelin Berthelot, 75005 Paris, France}

 \author{B. Muraz}
\affiliation{Laboratoire Kastler Brossel, Coll\`ege de France,
 CNRS, ENS-Universit\'e PSL,
 Sorbonne Universit\'e, \\11, place Marcelin Berthelot, 75005 Paris, France}

   \author{A. Couto}
\affiliation{Laboratoire Kastler Brossel, Coll\`ege de France,
 CNRS, ENS-Universit\'e PSL,
 Sorbonne Universit\'e, \\11, place Marcelin Berthelot, 75005 Paris, France}

 \author{J.-M. Raimond}
\affiliation{Laboratoire Kastler Brossel, Coll\`ege de France,
 CNRS, ENS-Universit\'e PSL,
 Sorbonne Universit\'e, \\11, place Marcelin Berthelot, 75005 Paris, France}

\author{M. Brune}
\affiliation{Laboratoire Kastler Brossel, Coll\`ege de France,
 CNRS, ENS-Universit\'e PSL,
 Sorbonne Universit\'e, \\11, place Marcelin Berthelot, 75005 Paris, France}

  \author{S. Gleyzes}
 \affiliation{Laboratoire Kastler Brossel, Coll\`ege de France,
 CNRS, ENS-Universit\'e PSL,
 Sorbonne Universit\'e, \\11, place Marcelin Berthelot, 75005 Paris, France}

\hyphenation{Ryd-berg sen-sing ma-ni-fold}

\begin{abstract}
Rydberg alkaline earth atoms are promising tools for quantum simulation and metrology. When one of the two valence electrons is promoted to long-lived circular states, the second valence electron can be optically manipulated without significant autoionization. We harness this feature to demonstrate laser slowing of a thermal atomic beam of circular strontium atoms. By driving the main ion core 422 nm wavelength resonance, we observe a velocity reduction of 50 m/s without significant autoionization. We also show that a superposition of circular states undergoes very weak decoherence during the cooling process, up to the scattering of more than thousand photons. This robustness opens new perspectives for quantum simulations over long timescales with circular atoms, while simultaneously cooling their motional state. It makes it possible to mitigate the harmful effects of unavoidable heating due to spin-motion coupling during a quantum simulation.

\end{abstract}

\date{\today}

\maketitle
                                                                                                                                                                                                                                                                                                                                                                                                                                                                                                                                                                                                                                                                                                                                                                                                                                                                                                                                                                                                                                                                                                                                                                                                                                                                                                                                                                                                                                                                                                                                                                                                                                                                                                                                                                                                                                                                                                                                                                                                                                                                                                                                                                                                                                                                                                                                                                                                                                                                                                                                                                                                                                                                                                                                                                                                                                                                                                                                                                                                                                                                                                                                                                                                                                                                                                                                                                                                                                                                                                                                                                                                                                                                                                                                                                                                                                                                                                                                                                                                                                                                                                                                                                                                                                                                                                                                                                                                                                                                                                                                                                                                                                                                                                                                                                                                                                                                                                                                                                                                                                                                                                                                                                                                                                                                                                                                                                                                                                                                                                                                                                                                                                                                                                                                                                                                                                                                                               
Rydberg atoms are ideal tools for the thriving domains of quantum metrology and quantum simulation~\cite{adams_rydberg_2020,browaeys_many-body_2020-1}.  Arrays with hundreds of low-angular-momentum, $\ell$, alkali Rydberg atoms, interacting through their strong dipole-dipole coupling,  achieve quantum simulations of interacting spin systems in regimes barely accessible to classical computers~\cite{ebadi_quantum_2021,scholl_quantum_2021}. Circular states \cite{hulet_rydberg_1983} (with maximum $\ell$) trapped in ponderomotive potentials \cite{cortinas_laser_2020,ravon_array_2023} promise to extend the timescale of the simulations to much larger values due to their extremely long lifetime \cite{nguyen_towards_2018}. 

Circular states of alkaline earth atoms additionally feature an ionic core, which can be used to optically manipulate the atoms \cite{meinert_indium_2020,muni_optical_2022,holzl_long-lived_2024}. Laser trapping of low-$\ell$ alkaline-earth-like Rydberg atoms using the non-resonant interaction with ionic core transitions has already been demonstrated \cite{wilson_trapped_2019}. However, these levels rapidly auto-ionize when the ionic core is excited \cite{Lochead_number-resolved_2013,Madjarov_high-fidelity_2020}. For high-$\ell$ Rydberg states, and in particular for a circular state, the auto-ionization is utterly slow \cite{poirier_autoionization_1988,Marin-Bujedo_autoionization_2023}, as demonstrated in the case of strontium~\cite{teixeira_preparation_2020}, allowing for ionic core excitation without atomic loss. This feature was used to coherently manipulate a superposition of circular levels with laser light \cite{muni_optical_2022}, taking advantage of the quadrupole interaction between the two valence electrons in doubly excited states. It opens the way to the use of the ionic core fluorescence for a non-destructive detection of the circular state. Core fluorescence could also be used to laser-cool circular atoms \cite{bouillon_direct_2024}, an essential tool to compensate the uncontrollable heating occurring during the quantum simulator operation.

Here, we demonstrate the slowing down of a beam of strontium circular Rydberg atoms using a laser resonant with one of the ionic core transitions. The auto-ionization of the doubly excited states is sufficiently slow to have a negligible effect over the experimental timescale. We reduce the atomic velocity by $\sim 50$ m/s, corresponding to the scattering of more that 5\,000 photons. Furthermore, we slow down a superposition of two circular states (the core scattering more than $\sim 1\,200$ photons in this case) without significant loss of its quantum coherence.

The experimental set-up is described in details elsewhere \cite{teixeira_preparation_2020}. A thermal beam of strontium atoms propagates along the $Ox$ axis [Fig. 1(a)]. These atoms are transferred into into the state $51c\,5s_{1/2}$, where one valence electron is excited in the circular state with principal quantum number 51, the other being left in the ground state of the ionic core, using a combination of three laser pulses, one microwave (MW) pulse and a short radio-frequency (rf) pulse with a well-defined polarization. The excitation takes place inside an electrode structure used to apply a static electric field that defines a vertical quantization axis $Oz$. Once in the circular Rydberg state, the atoms fly across this structure and finally reach a state-selective field-ionization detector. We set the field in the detector to match the ionization threshold of the $51c\,5s_{1/2}$ state and record the number of detected atoms as a function of their time of flight (TOF) $t_d$ between laser excitation and detection. Since the excitation lasers intersect the atomic beam at a 45$^\circ$ angle, they select by Doppler effect atoms with an average velocity $v^0_i$. 
\begin{figure}
 \centering
\includegraphics[width=\linewidth]{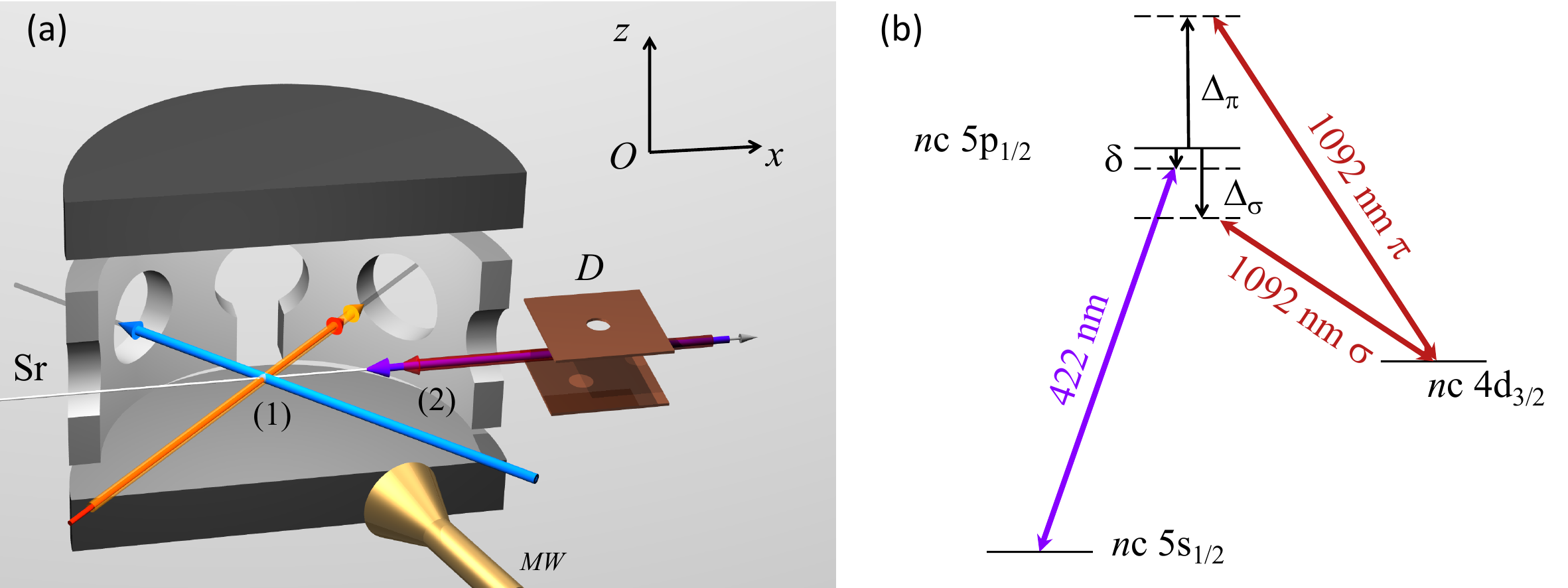}
 \caption{(a) Sketch of the experimental set-up. Strontium atoms (Sr) are excited to the circular state  $51c\,5s_{1/2}$ out of a thermal beam using a combination of lasers (461 nm, blue, 767 nm, orange, and 896 nm, red), MW and rf fields (for details see~\cite{teixeira_preparation_2020}). They then interact with the slowing (purple) and repumper lasers (dark red). We finally detect the atoms at the field-ionization detector $D$. A microwave source $MW$ allows us to prepare a superposition of two circular states.  (b) Energy levels and lasers involved during the ionic core laser excitation. The 422 nm wavelength laser drives the $nc5s_{1/2} - nc5p_{1/2}$ transition. The detunings $\Delta_{\pi,\sigma}$ correspond to the frequency difference between each repumper laser and the $nc4d_{3/2} - nc5p_{1/2}$ transition in the moving frame at velocity $v^0_i$.  }
 \label{fig:3}
\end{figure}
The preparation process is not perfect and leaves $\sim 15$\% of the atoms in other high-angular momentum states, close to the circular state, which have similar ionization thresholds and thus contribute to the ionization signal. We assume that they are slowed down exactly in the same way as the circular states, since their auto-ionization properties and ionic core level structure are very similar.  

Once the Rydberg electron is in a high-angular momentum state, the ionic core electron level structure is very similar to that of the Sr$^+$ ion [Fig. 1(b)]. In order to slow down the atoms, we shine a 422 nm slowing laser pulse, counter-propagating with the atomic beam, that drives the $51c\,5s_{1/2} - 51c\,5p_{1/2}$ transition. The power of this laser is 5.8 mW, corresponding to a measured Rabi frequency of about 30 MHz~\cite{Supplementarymaterial}. 
The laser frequency $\nu$ is chosen to be resonant with atoms at velocity $v^0_i$ at the beginning of the pulse. As the atoms are decelerated, the slowing laser frequency in the atomic frame decreases due to the Doppler effect by 2.4 MHz/(m/s)  \cite{Balykin_cooling_1980}. The laser frequency is thus increased linearly  during the pulse: $\nu(t)=\nu(0)+\alpha t$. The optimal slowing is observed for $\alpha\approx 750$~kHz/\mmu s.

The $51c\,5p_{1/2}$ state decays into the ionic ground state and into the $51c\,4d_{3/2}$  ionic metastable state (branching ratio 1:17 \cite{zhang_iterative_2016}).  To recycle the atoms trapped in the latter, we drive the $51c\,4d_{3/2} - 51c\,5p_{1/2}$ transition (frequency $\nu_{1092}$ in the moving frame at velocity $v_i^0$) using two repumper beams with orthogonal polarizations ($\sigma$ and $\pi$). These two beams are overlapped with the 422~nm beam.
 They must have different frequencies $\nu_\sigma$  and $ \nu_\pi = \nu_\sigma + \Delta$, in order to prevent the appearance of a dark state on the $51c\,4d_{3/2}-51c\,5p_{1/2}$ transition~\cite{berkeland_destabilization_2002}. The  detuning $\Delta_\sigma$ and $\Delta_\pi$ with respect to $\nu_{1092}$ [Fig. 1(b)] also need to be both sufficiently large to ensure that the slowing and repumper laser frequencies do not match the  Electromagnetically Induced Transparency (EIT) condition that would make the atom impervious to the slowing light \cite{Boller_observation_1991}. 
 
 We operate in a regime where $\Delta$ is large ($\Delta = 100$ MHz) and choose $\Delta_\sigma \approx -25$ MHz and $\Delta_\pi  \approx 75$ MHz.
 Each repumper laser has a power of $\sim 10$~mW, corresponding to a Rabi frequency of $\sim20$~MHz~\cite{Supplementarymaterial}, 
large enough compared to the Doppler shift variation for the repumper lasers during the experiment ($\sim 60$ MHz) 
that we do not need to ramp $\nu_{rep}$. The chosen detunings also ensures that we never match any of the two EIT resonance conditions as the atom slows down.

\begin{figure}
 \centering
 \includegraphics[width=\linewidth]{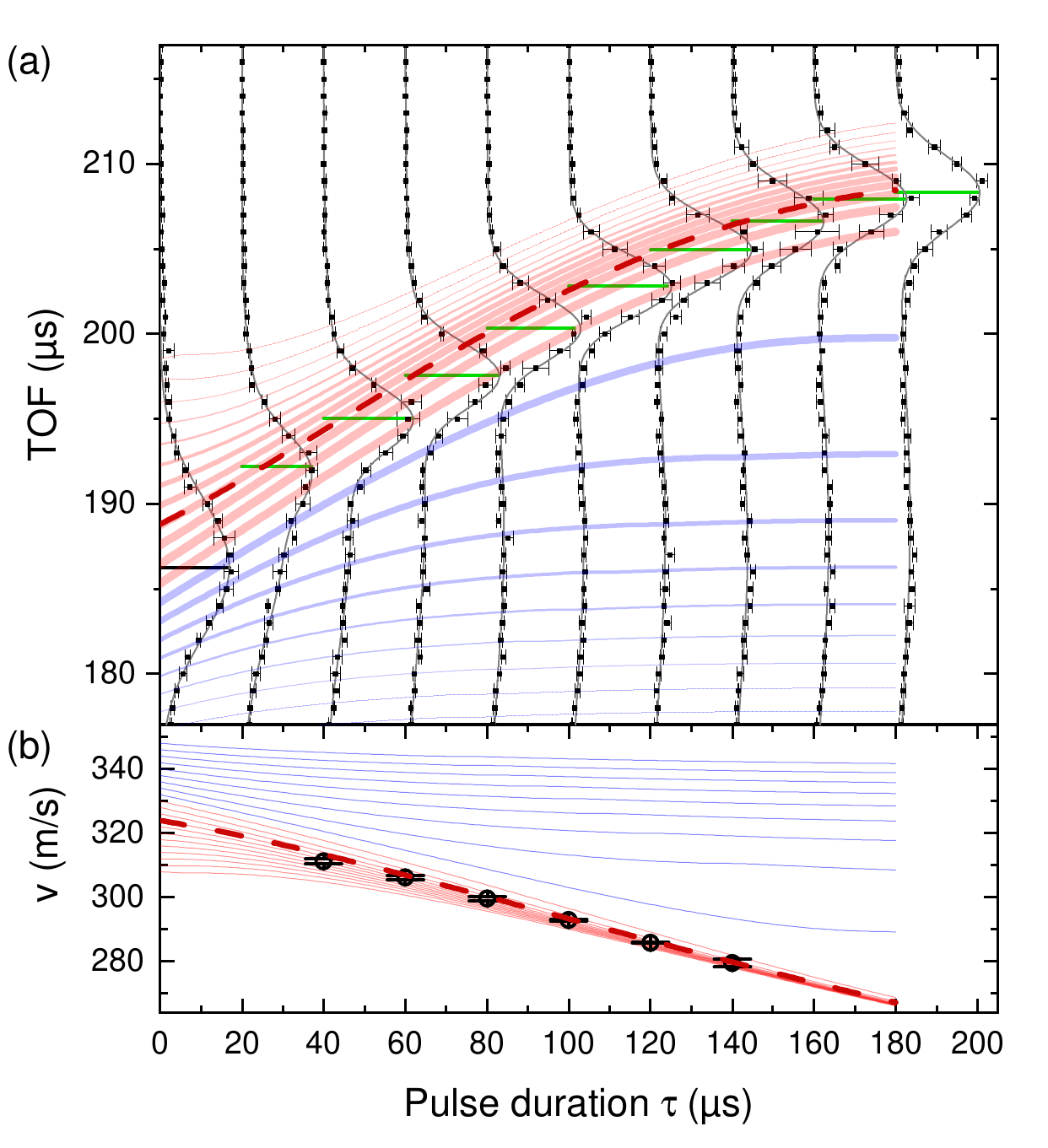}
 \caption{(a) Time of flight signal as a function of the slowing laser pulse duration $\tau$. The black points are the experimental data (with statistical error bars), the black lines the result of a fit  to a gaussian peak (for $\tau = 0$) or to a sum of two gaussian peaks (for $\tau >0$). The thick black horizontal bar marks the TOF peak position $t_0$ without slowing laser. The green horizontal bars mark the center positions $\bar t_{d}$ of the gaussian peaks corresponding to the slowed atoms for $\tau >0$. The colored lines are the results of numerical simulations computing, for different initial velocities $v^\prime_i$, the TOF of the atoms as a function of the slowing laser pulse duration. The thickness of the line is proportional to the probability for the atom to have the initial velocity $v^\prime_i$. The color of the line indicate whether the atoms are in the 2/3 slowest (red) or 1/3 fastest (blue) fraction of the initial velocity distribution. The dashed red line shows the average TOF of the slowest atoms. (b) Black open dots are the measured atomic velocities $v(\tau)$.  Blue and red solid lines show the result of a numerical simulation of the atomic velocity variation for different initial velocities $v^\prime_i$ as in (a). The dashed red line shows the average velocity of the slowest atoms computed from the simulations.}
 \label{fig:2}
\end{figure}

Figure 2(a) shows the TOF distribution as a function of the slowing laser pulse duration $\tau$. For $\tau = 0$ (no slowing laser), the distribution is close to a gaussian centered on $t_0= 186.3$ \mmu s with a $4.2\ \mmu$s width. From the experimental set-up geometry, we deduce the average initial velocity $v^0_i = 328.2$~m/s and its dispersion $\Delta v_i = 7.2$~m/s.  
As $\tau$
increases, we observe that the main peak in the TOF distribution occurs at a later time, clearly demonstrating the slowing. 

The red and blue lines on Fig. 2(a) represent the numerical simulation of the TOF evolution as a function of $\tau$ for different initial atomic velocities (see \cite{Supplementarymaterial}). 
Atoms with initial velocities $v_i \lesssim v^0_i$ (red lines) are slowed down by the laser. Atoms with $v_i \approx v^0_i$ are resonant from the start of the pulse, atoms with lower velocities ($v_i < v^0_i$) eventually come into resonance because of the slowing laser frequency sweep. This leads to a reduction of the velocity dispersion of the slow atom packet, which explains why the main peak in the TOF distribution becomes narrower for larger $\tau$ [Fig 2(a)].
The fastest atoms ($v_i >  v^0_i$, blue lines) are too fast to be efficiently slowed. They rapidly stop interacting with the slowing laser and form a broad second peak in the TOF distribution, with a position nearly independent of $\tau$. 
For each duration $\tau > 0$, we fit the TOF distribution to the sum of two gaussian peaks [black solid lines on Fig. 2(a)].  From the relative area of these peaks, we determine that $58\%$ of the atoms are slowed down. The green horizontal lines [Fig. 2(a)] mark, for each $\tau$ value, the mean value $\bar t_{d}$ of the TOF of the slowed atoms.

To estimate the atomic deceleration, we measure the average velocity $v (\tau)$ of the slowed atom packet as a function of the slowing laser pulse duration. For each value of $\tau$ with $40 \le \tau\le 140$~$\mu$s, we vary the pulse start time $t_s$. The atoms thus fly at a velocity $v(0)$ between $t=0$ and $t=t_s$, and at a velocity $v(\tau)$ between $t=t_s+\tau$ and their detection. Therefore, the TOF $\bar t_{d}$ of the slowed atoms varies linearly with $t_s$, with a slope that measures $v (\tau)/v(0)$~\cite{Supplementarymaterial}. We can thus reconstruct $v (\tau)$ [black points on Fig. 2(b)]. The results are consistent with a constant deceleration and in excellent agreement with a numerical simulation (dashed red line of Fig. 2(b)). A fit to the data gives a deceleration $a=0.33(1)$~m/s/\mmu s, which corresponds to the scattering of 30.9(9) photons per microsecond. Given the decay rate of $51c\,5p_{1/2}$ ($\Gamma^{-1} = 7.4$~ns), 
this means that the $51c\,5p_{1/2}$p state population is 22.9 \%.

The doubly excited $51c\,5p_{1/2}$ state is, in principle, prone to autoionization. Scattering the 422 nm photons could thus lead to the loss of the Rydberg atoms. To set a limit on the autoionization rate, we measure the total number of detected atoms as a function of the slowing laser pulse duration. We observe a small decrease of the atom number, corresponding to a fractional loss of $1.5(6)\cdot 10^{-4}$ per \mmu s. This marginal loss could be caused by many experimental imperfections (for instance, the atoms could merely be deflected away from the center of the detector by the laser radiation pressure). Assuming as a worst case scenario that this loss is entirely due to the auto-ionization of $51c\,5p_{1/2}$, and taking into account that $58\%$ of the atoms interact with the laser and are thus in $51c\,5p_{1/2}$ 22.9\% of the time, we get an upper limit of the auto-ionization rate of $1.1(5)$~kHz corresponding to an auto-ionization probability of less than $10^{-5}$ per scattered photon. 

Next, we study the effect of the slowing laser on the coherence of a quantum superposition of the two circular states $51c$ and $50c$. In the presence of the slowing and repumper lasers, the ionic core electron randomly jumps between the $5s_{1/2}$, $5p_{1/2}$ and $4d_{3/2}$ levels. Switching from  $5s_{1/2}$ to $5p_{1/2}$ has a negligible effect on the $50c-51c$ transition frequency. However, when the electron is in one of the $4d_{3/2},m_j$ states, the residual Coulomb interaction between the two valence electrons induces a small shift $\pm \Delta\nu/2$  ($\Delta\nu = 95$~kHz) of the $50c-51c$ transition frequency whose sign depends on $|m_j|$~\cite{muni_optical_2022}. As a result, the Rydberg state superposition accumulates a stochastic phase $\Phi$ that depends on the relative time spent in the different $4d_{3/2},m_j$ sublevels and affects its quantum coherence.

We investigate this decoherence by Ramsey spectroscopy. We apply onto the $51c$ atoms two $\pi/2$ MW pulses at 51.0985 GHz (duration 2.2 \mmu s and 1.4 \mmu s respectively), resonant with the $50c-51c$ transition and separated by a time interval $T=40$~\mmu s, limited by the MW mode structure in the experiment. To ensure that a stationary regime has been reached for the ionic core state, the first MW pulse  occurs 3 \mmu s after the beginning of the slowing laser pulse, which stops after a duration $\tau = 45 \ \mmu$s, right after the second MW pulse.
Finally, we record the number of atoms detected in $50c$ as a function of the relative phase $\varphi_R$ between the two MW pulses.  During the Ramsey sequence, the atoms scatter 1200 photons and their velocity changes by about 13 m/s.
Note that non-circular states in the 51 manifold are not affected by the MW pulses and do not contribute to the signal. Detection of $50c$ effectively post-selects the atoms correctly prepared in $51c$. 

\begin{figure}
 \centering
  \includegraphics[width=.9\linewidth]{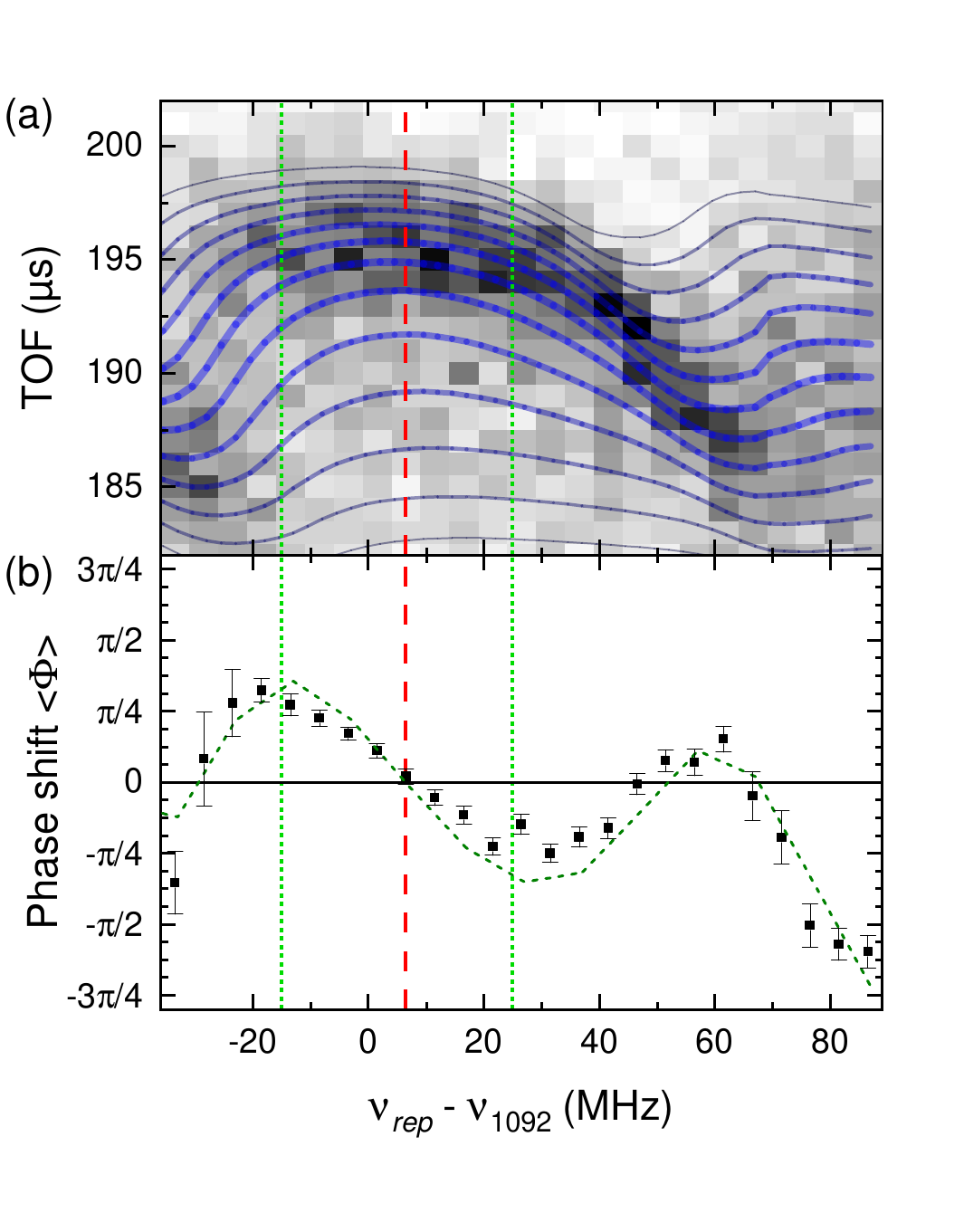}
 \caption{(a) Time of flight signal for a $\tau = 45 \mu$s slowing laser pulse as a function of the {relative frequency $\nu_{rep} - \nu_{1092}$}. The grayscale plot shows the experimental data (gray level scale normalized to the highest count), the blue lines are the result of a numerical simulation of the TOF for different initial velocity $v^\prime_i = v_i-12$~m/s, $v_i-10$~m/s, ..., $v_i+12$~m/s. The thickness of each line is proportional to the probability for an atom to have the initial velocity $v^\prime_i$. (b) Phase shift of the Ramsey fringes induced by the slowing laser as a function of the relative repumper frequency. The black points are the experimental measurements (with statistical error bars). The green dashed line is the result of a simple model assuming $\Phi^{th} = 2 \pi T \,\Delta\nu [\langle \pi_{|m_j|=3/2} \rangle -\langle \pi_{|m_j|=1/2} \rangle]/2$. }
 \label{fig:4}
\end{figure}

We first check the slowing efficiency as a function of the repumper beams detuning by scanning simultaneously $\nu_\pi$ and $\nu_\sigma$, keeping $\Delta$ constant. Fig \ref{fig:4}(a) presents the gray scale map of the experimental $51c$ TOFs as a function of $\nu_{rep}=( \nu_\pi+\nu_\sigma)/2$ when no MW pulses are applied. We observe that the repumper frequencies can be varied over a range of $\sim 40$ MHz without noticeably affecting the slowing efficiency (area limited by green dotted lines). Outside this range, one of the repumper beams gets too close to the EIT condition, resulting in an inefficient slowing. For different values of $\nu_{rep}$, we  record the Ramsey fringes. Fig. \ref{fig:4}(b) presents the measured fringes phase shift $\langle \Phi \rangle$ (black squares) obtained by a sine fit of $50c$ counts.
The average accumulated phase is very sensitive to the repumper laser frequency, which controls the amount of time spent by the ionic core electron in the $4d_{3/2},m_j$ sublevels.
The values of $\langle \Phi \rangle$ are in very good agreement with a shift proportional to the population difference $\langle \pi_{|m_j|=3/2} \rangle -\langle \pi_{|m_j|=1/2} \rangle$ of the $|m_j|$ sublevels of the $4d_{3/2}$ ionic core state, state, obtained by a numerical simulation and averaged over the duration $T$ of the Ramsey sequence (green dashed line).
The repumper frequency can be chosen so that $\langle \Phi \rangle = 0$ (red dashed vertical line on Fig. \ref{fig:4}(b)).

Ultimately, the stochastic nature of $\Phi$ limits the coherence of the Rydberg superposition. To observe this decoherence, we set the repumper frequency to cancel $\langle \Phi \rangle$ and measure the Ramsey fringes visibility with and without the slowing laser pulse. Figure 4 shows the variation of the number $N$ of atoms detected in $50c$ as a fonction of $\varphi_R$. To limit the contribution of other experimental sources of decoherence (such as the electric field inhomogeneity along the atomic path through the apparatus), we only consider the atoms detected in a narrow 3 \mmu s time-window around the center of TOF distribution \cite{Supplementarymaterial}. From a sine fit to the data, we get a fringe visibility of 91\% when no laser is applied. The visibility decreases to 85\% in the presence of the slowing laser. A $6\%$ visibility loss  corresponds to a coherence loss smaller than $10^{-4}$ per scattered photon, consistent with the numerical simulation results \cite{Supplementarymaterial}.

\begin{figure}
 \centering
 \includegraphics[width=.95 \linewidth]{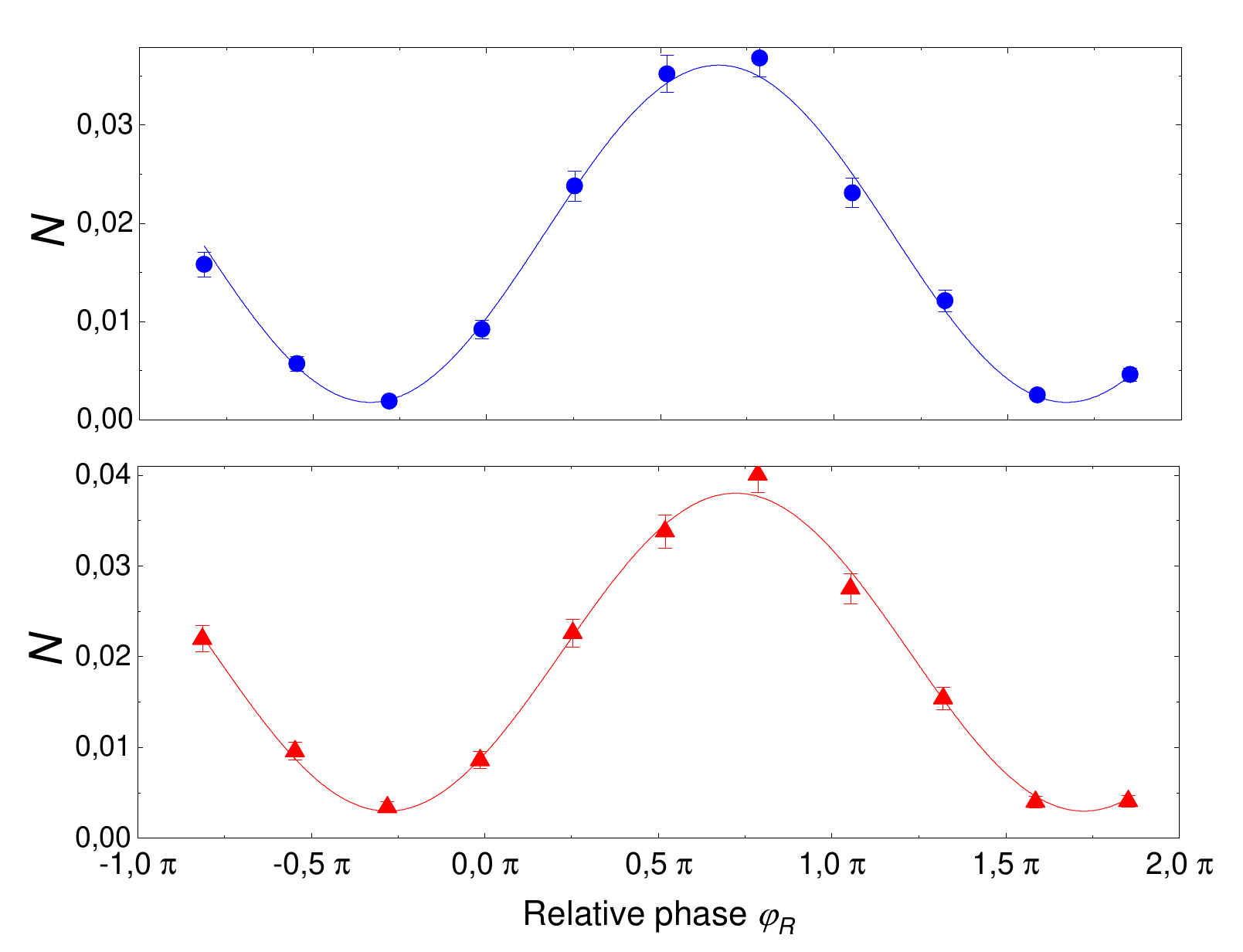}
   \caption{(a) Ramsey fringes in the absence of cooling laser.
 Number of atoms detected in $50c$ in a 3 \mmu s time window around $ t_d =186 $~\mmu s as a function of the relative phase $\varphi_R$ between the MW pulses. The points are the experimental data, the solid line a sine fit. (b)  Number of atoms detected in $50c$ in a 3 \mmu s time window around $ t_d =195 $~\mmu s  [red line on Fig. 4(b)] in the presence of the slowing laser, when the frequencies of the 1092~nm repumpers are set so that $\langle \Phi \rangle = 0$.   The points are the experimental data, the solid line a sine fit.}
 \label{fig:3}
\end{figure}

We have shown that it is possible to slow down a beam of strontium circular Rydberg atoms by applying a counter-propagating laser resonant with {the main} ionic core optical transition. {The observed velocity variation $\Delta v \approx 50$~m/s is only limited by the distance separating the atom preparation and detection zones. During the 180$\ \mu$s slowing laser pulse duration, circular atoms scatter about 5000 photons, without appreciable auto-ionization. This opens the way to the use of sub-doppler cooling methods developed in the context of trapped ions for laser-cooling the residual motion of circular atoms trapped in optical tweezers.}

We have also measured the weak decoherence induced by the slowing beam to a circular state superposition due to the residual electrostatic coupling between the two valence electrons. 
This  decoherence could be further reduced by dynamical decoupling techniques. One could simply apply a strong microwave drive on the $51c -50c$ transition, so that the new eigenstates of the system [$ (\ket {51c} \pm i \ket {50c})/\sqrt 2$], have the same electrostatic shift regardless of the ionic core state. 

The robustness of circular state coherence during laser cooling opens interesting perspectives for applications of strontium circular Rydberg states to quantum technologies. Laser cooling of  trapped circular Rydberg atoms in optical tweezers could be essential for a better localization of the atoms in quantum metrology and quantum simulation experiments~\cite{}. It would freeze any residual motion of the  atoms in the array before starting the experiment or compensate for the heating during the experiment. This is an important asset for exploiting the long lifetime of circular atoms for long time scale quantum simulations or for high precision quantum metrology.

\begin{acknowledgments}

The authors would like to thank I. Dotsenko for experimental support. 
This publication has received funding under Horizon 2020 program via the project 786919 (Trenscrybe) and under Horizon Europe programme HORIZON-CL4-2022-QUANTUM-02-SGA via the project 101113690 (PASQuanS2.1). Data are available upon reasonable request.

\end{acknowledgments}

\renewcommand{\theequation}{S\arabic{equation}}
\renewcommand{\thefigure}{S\arabic{figure}}
\setcounter{figure}{0}  

\clearpage
\section{Supplementary Information}
We describe in more details the velocity measurement, the effect of the electric field inhomogeneity, the numerical model and the parameter estimation. 

\subsection{Velocity measurement}

To determine the average velocity $v(\tau)$ of the slowed atom packet, we record the TOF signal for different durations $\tau$ and different start times $t_s$ of the slowing pulse. For each value of $\tau$, the atom packet flies at the initial average velocity $v(0)$ between $t=0$ (preparation of the circular atom) and $t=t_s$. During the laser pulse, it is decelerated to the velocity $v(\tau)$ over a distance $d(\tau)$. From $t=t_s+\tau$ to the time $\bar t_{d}$, at which it reaches the detector, it flies at a constant velocity $v(\tau)$. We thus have 
\begin{equation} 
v(0) t_s + d(\tau) + (\bar t_{d} - t_s - \tau) v(\tau) = D
\label{vtau}
\end{equation} 
where $D$ is the distance between the laser excitation and the detector. Solving (\ref{vtau}), we get
$$ \bar t_{d} = \frac{\Delta v(\tau) /v(0)}{1+\Delta v(\tau)/v(0)} t_s + \frac{D-d(\tau)}{v(\tau)} + \tau $$
where $\Delta v(\tau) = v(\tau) - v(0) $. {For} each value of $\tau$, we {measure $\bar t_{d} (t_s)$ and perform a linear fit of its variation with $t_s$} [Fig. S1(a)]. Fig. S1(b) presents the values of $ \Delta v(\tau) /v(0)$ deduced from the slope of the fits. We observe that $\Delta v(\tau)$ is proportional to $\tau$, meaning that the laser induces a constant deceleration $a_0$. To determine $v(0)$, we fit again the data $ \bar t_{d} (\tau, t_s)$, assuming a constant deceleration $a = \eta v(0)$ [where $\eta$ is the slope of the fit in Fig S1(b)]. We get $v(0) = 325.6\pm 0.06$~m/s and  $a = 0.3302\pm 0.0007 $~m/s/\mmu s. Note that $v(0)$ is smaller than $v^0_i$ as only the slowest atoms are decelerated by the laser.

A direct fit of the $ \bar t_{d}$  data from Fig. 2(a) of the main text to Equ.(\ref{vtau}), assuming a constant deceleration, gives $v(0) = 324.9\pm 0.24$~m/s and  $a = 0.318\pm 0.003 $~m/s/\mmu s. The comparison with the previous value provides an estimate of the uncertainty in the determination of $a$.

\begin{figure}
 \centering
 (a)\hspace{.85\linewidth} \; \\ \vspace{-10mm}
 \includegraphics[width=.95 \linewidth]{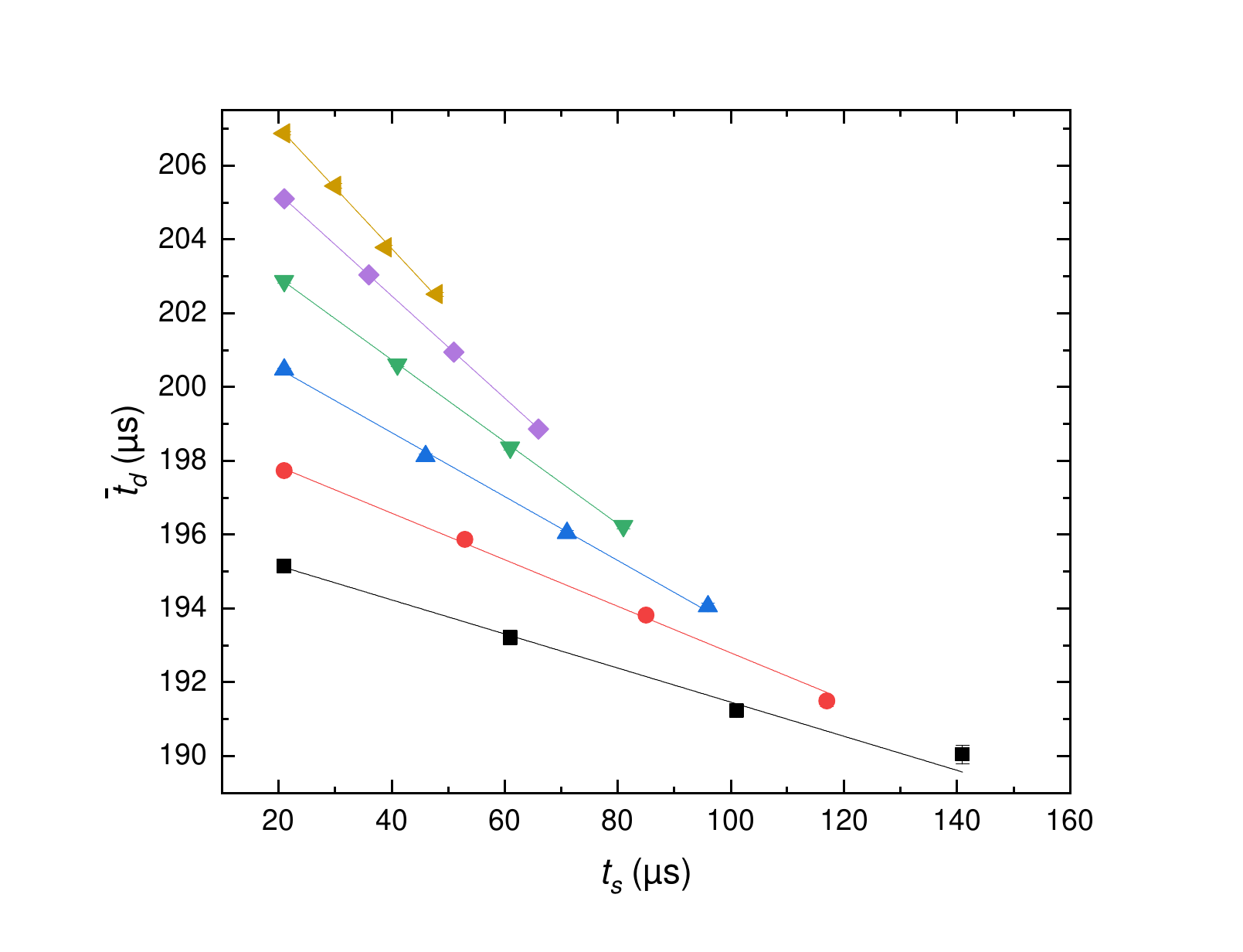}
 (b)\hspace{.85\linewidth} \; \\ \vspace{-10mm}

  \includegraphics[width=.95 \linewidth]{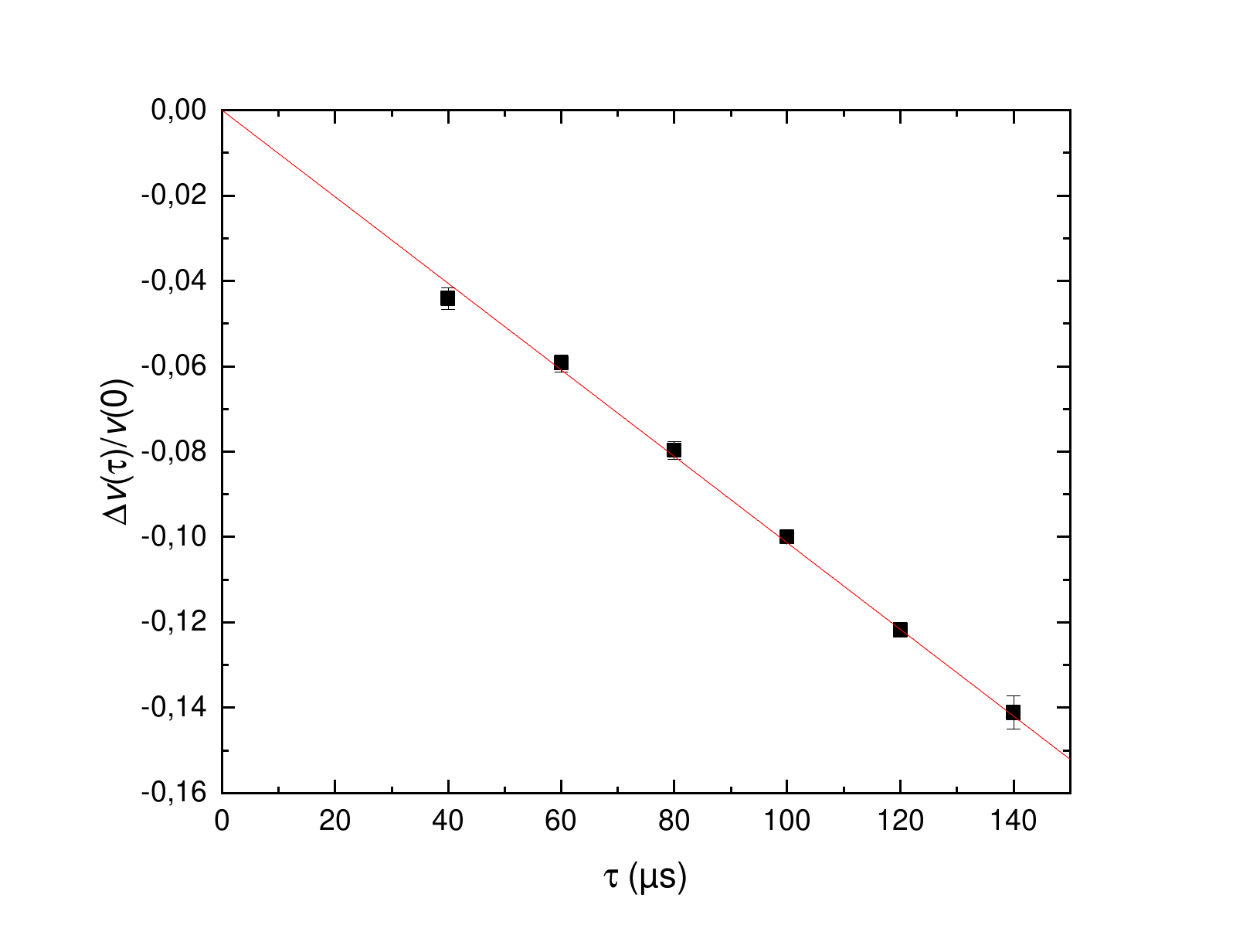}
   \caption{ (a) Time of flight $\bar t_{d}$ of the slowed atom packet as a function of the start time of the slowing pulse $t_s$ for different values of $\tau$ : 40 \mmu s (black), 60 \mmu s (red), 80 \mmu s (blue), 100 \mmu s (green), 120 \mmu s (purple), 140 \mmu s (yellow). The points are experimental (with statistical error bars\comment{on ne les voit pas}), the lines are the result of a linear fit of the data. (b) Relative deceleration $\Delta v(\tau) /v(0)$ as a function of $\tau$. The points are the values deduced from the fit shown in (a), the solid line a fit to the equation $\Delta v(\tau) /v(0) = \eta \tau$ from which we obtain $\eta = 1.014(2)\cdot10^{-3}$. }
\end{figure}


\subsection{Inhomogeneity effect}
To estimate the decoherence induced by the lasers on the circular state superposition, we record the number $N(t_d,\varphi_R)$ of atoms detected in $50c$  as a function of the TOF $t_d$ and the relative phase $\varphi_R$. Fig. S2 (central panel) presents the experimental data without (a) and with (b) the slowing laser pulse. 
 For each TOF $t_d$, we fit $N(t_d,\varphi_R)$ to
$$ \frac 1 2 \left( N_0(t_d) + A(t_d) \cos \left(\varphi_R - \varphi(t_d)\right) \right).$$
The fit parameter $N_0(t_d)$ directly gives the TOF distribution (left panels of Fig. S2). The ratio $A(t_d)/N_0(t_d)$ measures the visibility of the fringes (and hence the degree of coherence of the superposition) for each TOF $t_d$ (right panels of Fig. S2). The green line on the right panels shows the averaged visibility over the whole atomic packet (weighted by $N_0(t_d)$). The red line shows the visibility of the fringes averaged over the whole atomic packet:
$$N(\phi)=\sum_{t_d} N(t_d,\varphi_R). $$
 
 \begin{figure}
 \centering
 \includegraphics[width=.95 \linewidth]{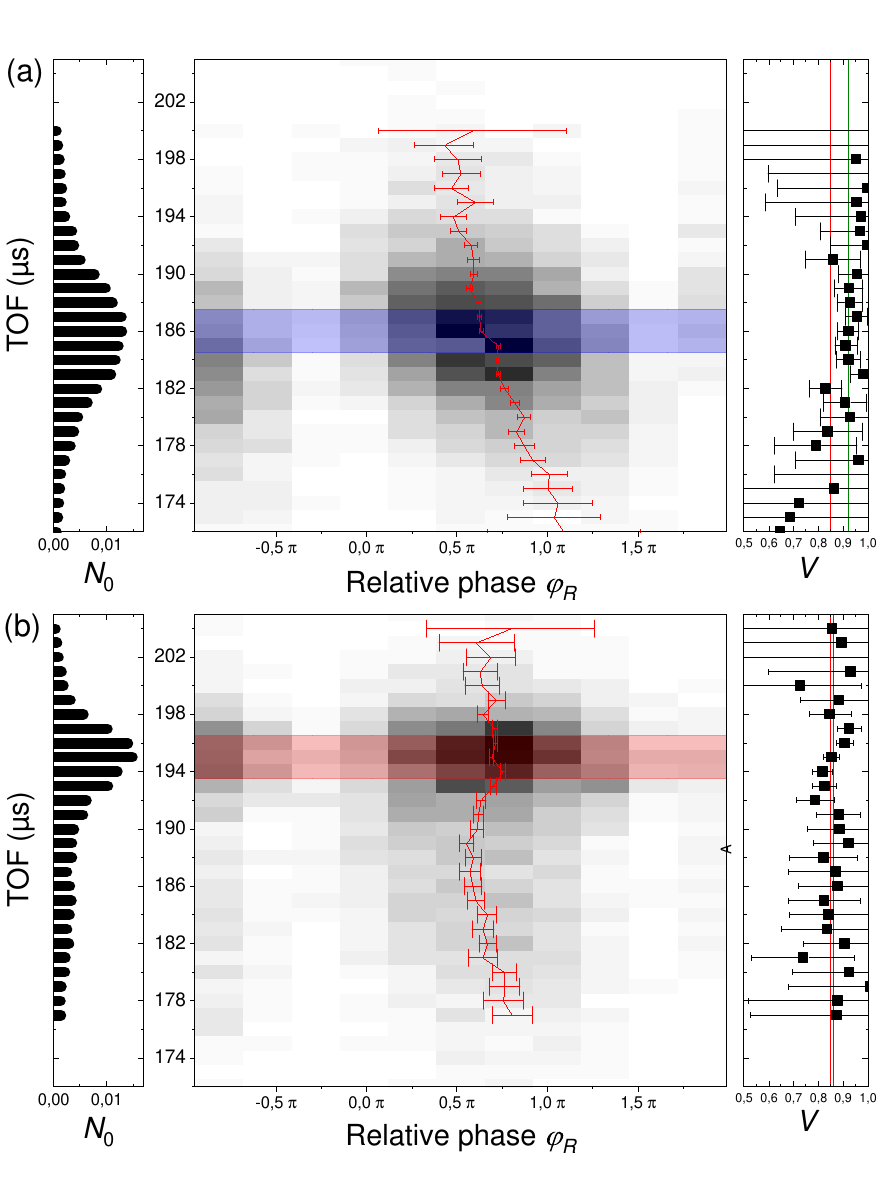}
   \caption{(a) {Center panel : the grayscale plot shows $N(t_d,\varphi_R)$ in the absence of slowing laser. The red solid line shows the value of $\varphi(t_d)$ (with statistical error bars from the fit). Left panel $N_0(t_d)$. Right panel : Visibility $A(t_d)/N_0(t_d)$ of the fringes for each TOF $t_d$ (with error bars deduced from the fit errors). The green line is the average of the visibility (weighted by $N_0(t_d)$), the red line is the visibility of the fringes averaged over the atomic packet. (b) Same with the slowing laser applied.} The colored area correspond to the time window used to compute the data of Fig. 4 of the main text.}
\end{figure}
One clearly sees that, in the absence of laser, the visibility of the averaged fringes is much lower than the average of the visibility.  We estimate that this effect is due to the residual gradient of the static electric field applied on the atoms. It causes a variation of the phase $\varphi(t_d)$  with $t_d$ (red lines on the central panels). To limit our sensitivity to this effect, we only consider, for the data presented in Fig. 4 of the main text, a narrow 3 \mmu s  time window around the center of the TOF peak {(Blue and red shaded area on figure S2)}. 

\subsection{Numerical model}

In order to model the optical pumping dynamics induced by the slowing and repumper lasers, we describe the ionic core internal state by a density matrix $\rho$ taking into account the 8 levels $ \ket {51c\,5s_{1/2},m_j} $,  $ \ket {51c\,5p_{1/2},m_j} $ and $ \ket {51c\,4d_{3/2},m_j} $. We describe the atomic motion by the atom position $x(t)$ along the atomic beam axis ($x=0$ corresponds to the laser excitation, and $x=D$ to the position of the detector). 

We numerically solve a coupled system with, on the one hand, the Lindblad master equation  (in which the Hamiltonian H depends on $v(t)$ through the Doppler effect of the slowing and repumper lasers) and, on the other hand, the equations of motion 
$$ \dot x (t) = v(t) $$
$$ \dot v(t) = - a_\mathrm{max} \sum_{m_j=-1/2}^{1/2} \rho_{\ket {51c\,5p_{\frac 1 2},m_j},\ket {51c\,5p_{\frac 1 2},m_j}}$$
where $a_\mathrm{max}\approx 1.44$~m/s/\mmu s is the acceleration corresponding to one 422 nm photon recoil per $5p_{1/2}$ lifetime (7.4 ns).

We choose for the initial conditions a statistical mixture with equal weights of the two $\ket {51c\,5s_{1/2},m_j} $ levels, $x(0)=0$ and $v(0)=v^\prime_i$ with $v^\prime_i = v_0 - 20$~m/s, $v_0 - 18$~m/s, ...,  $v_0 + 20 $~m/s. \comment{qu'est ce que v0. on l'a appellee$ v_i $dans le main text non?} To compute {$\langle \pi_{|m_j|=3/2} \rangle -\langle \pi_{|m_j|=1/2} \rangle$, we sum the contribution of each} $v^\prime_i$ with a weight $\propto \exp[-2(v^\prime_i)^2/(w)^2]$, where $w = 14.7$ m/s.

\subsection{Parameters estimation}

To estimate the intensity of the slowing laser at the position of the atoms, we measure the Autler-Townes doubletit produces on the $ {51c\,5s_{1/2}} -  {51c\,5p_{1/2}} $ transition by using the repumper laser as a probe.

To estimate the intensity of the repumper laser beams at the position of the atoms, we detune them from the $ {51c\,4d_{3/2}} -  {51c\,5p_{1/2}} $ transition and set $\Delta$ to induce a resonant Raman transfer between the $|m_j|=1/2$ and $|m_j|=3/2$ sublevels of $ {51c\,4d_{3/2}} $ as described in \cite{muni_optical_2022}. The Rabi frequency of the Raman transfer allows to measure the product $\Omega_\sigma \Omega_\pi$, where $\Omega_\sigma$ (resp. $\Omega_\pi$) is the Rabi frequency proportional to the amplitude of the $\sigma$ (resp. $\pi$) component of the laser field. Since the two beams with orthogonal polarizations are sent through the same single mode fiber, the ratio $\Omega_\pi^2/\Omega_\sigma^2$ is given by the ratio of the two beam powers measured at the output of this fiber, allowing us to deduce the values of $\Omega_\sigma$ and $\Omega_\pi$.

We assume that, at the beginning of the slowing pulse ($\tau = 0$), the slowing laser frequency is resonant with the atoms that have the central velocity $v^0_i$. This is confirmed by the results of the simulation : changing the detuning of the slowing laser changes the population balance between the two peaks in the TOF distribution. The ratio that we obtain in the simulation (2/3 of the atoms are slowed down by the laser -- see red lines in Fig. 2 of the main text) is in very good agreement with the ratio we measure (58 \% of the atom in the second peak of the TOF).

\subsection{Laser induced decoherence model}
 
 To simulate the effect of the slowing laser on the coherence of a superposition of two circular states, we now calculate the evolution of the density matrix taking into account the levels $ \ket {nc,5s_{1/2},m_j} $,  $ \ket {nc,5p_{1/2},m_j} $ and $ \ket {nc,4d_{3/2},m_j} $ for both $n=51$ and $n=50$. We take into account the $|m_j|$ dependent frequency shift $\Delta \nu=  {95}$~kHz of the $50c$ to $51c$ transition when the ionic core is in $4d_{3/2}$ for computing the unitary part of the evolution. We assume that the jump operators describing the decay from $ nc5p_{1/2}$ into $nc4d_{3/2}$ $m_j$ sublevels preserve the coherence of the circular state superposition. Physically, it means that the frequency of the photons emitted when the ionic core decays from  $  {5p_{1/2}} $ to $  {4d_{3/2}} $ does not depends sufficiently on $n$ to provide a which-path information as whether the atom is in $50c$ or $51c$.
 
 Fig. S3 shows the evolution of the Rydberg coherence {defined as the visibility of fringes} as a function of the number of scattered 422 nm photons, for different values of the initial velocity, in the conditions of Fig. 4, for a 45~\mmu s slowing laser pulse. We observe that the efficiency of this pulse depends on the initial velocity (as the Doppler effect changes the frequency of the slowing and repumper lasers seen by the atom). The observed coherence loss remains on the order of $10^{-4}$ per scattered photon. 
 
  \begin{figure}
 \centering
 \includegraphics[width=.95 \linewidth]{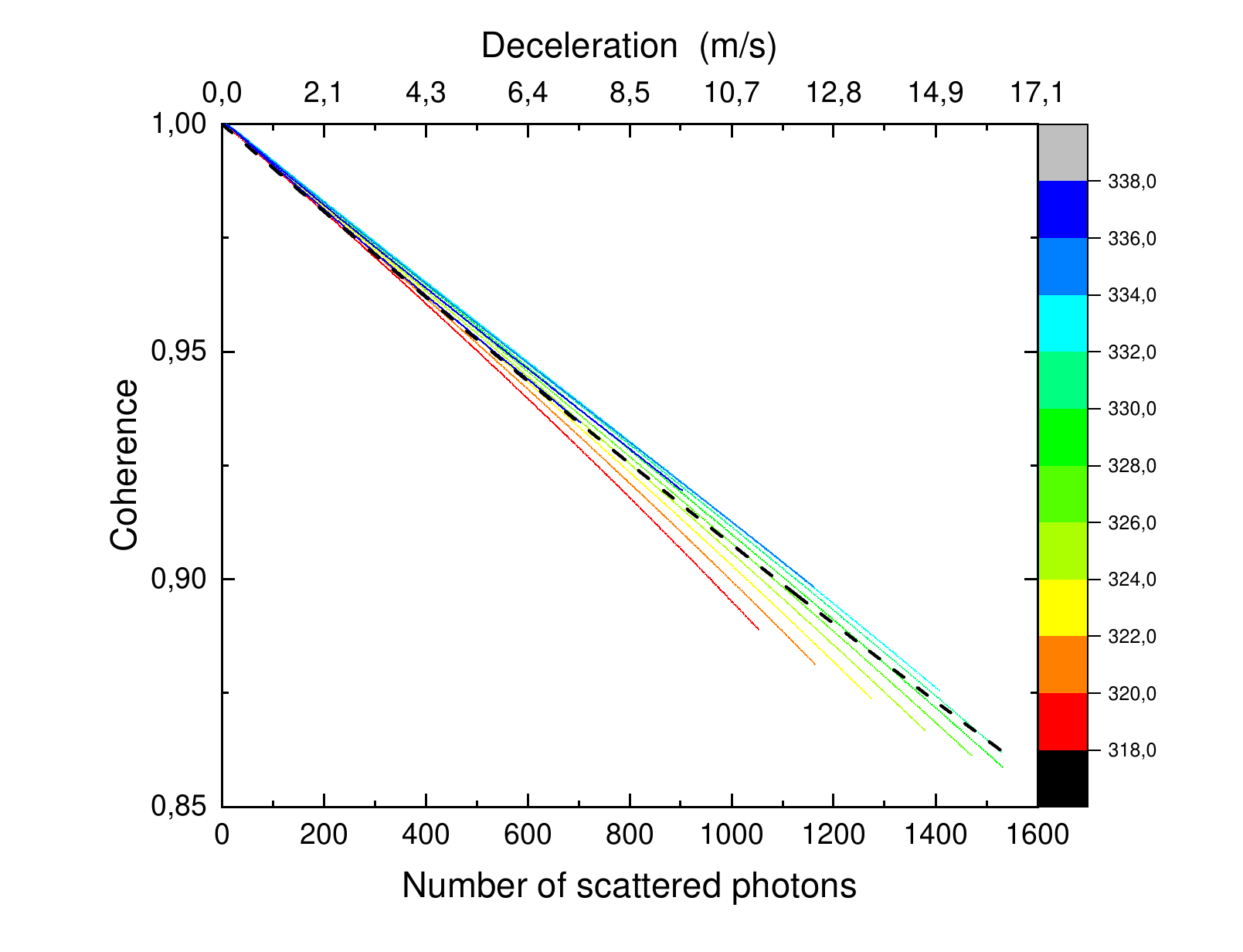}
   \caption{Coherence of the circular state superposition during a 45 \mmu s slowing pulse as function of the number of scattered photons for different initial velocities (solid line, color code for the velocity on the right). The coherence typically decay with a rate of $10^{-4}$ per scattered photon.}
\end{figure}

\end{document}